\documentclass[journal]{IEEEtran}

\ifCLASSINFOpdf
\else
\fi

\usepackage{hyperref}        
\usepackage{url}             
\usepackage{booktabs} 
\usepackage{amsfonts} 
\usepackage{nicefrac}        
\usepackage{microtype}       
\usepackage{bookmark}
\usepackage{array}
\usepackage{graphicx}
\usepackage{amsmath}
\usepackage{amsthm}
\usepackage{amsfonts} 
\usepackage{amssymb}
\usepackage{array}
\usepackage{algorithmic}
\usepackage{algorithm}
\usepackage{tabularx}
\usepackage{subfigure}
\usepackage{color}
\usepackage{enumitem}
\usepackage{cite}

\usepackage[T1]{fontenc}
\usepackage{amsmath}
\usepackage[cmintegrals]{newtxmath}
\usepackage{bm}

\usepackage{multirow}

\hypersetup{hidelinks}

\newtheorem{remark}{Remark}

\definecolor{airforceblue}{rgb}{0.36, 0.54, 0.66}
\definecolor{applegreen}{rgb}{0.55, 0.71, 0.0}
\definecolor{bittersweet}{rgb}{1.0, 0.44, 0.37}

\LetLtxMacro{\originaleqref}{\eqref}

\renewcommand{\eqref}{Eq.~\originaleqref}
\newcommand{\refcite}{Ref.~\cite}

\newcommand{\rv}{\color{black}}

\title{\Huge Conservative Sparse Neural Network Embedded Frequency-Constrained Unit Commitment \\ 
With Distributed Energy Resources}

\author{
  Linwei~Sang,
  Yinliang~Xu$^*$,~\IEEEmembership{Senior Member,~IEEE,}
  Zhongkai~Yi,
  {Lun~Yang,}
  Huan~Long,~\IEEEmembership{Member,~IEEE,}
  and Hongbin~Sun,~\IEEEmembership{Fellow,~IEEE}
  \thanks{Manuscript received Jul.~15, 2022; revised Mar.~6, 2023; accepted Apr. 14, 2023. This work was supported by the by Shenzhen Science and Technology Program, Grant No. WDZC20220808143010001, and National Natural Science Foundation of China, Grant No. 52277107. (Corresponding Author: Yinliang Xu).}
  \thanks{Linwei~Sang and Yinliang~Xu are with Tsinghua-Berkeley Shenzhen Institute, Tsinghua Shenzhen International Graduate School, Tsinghua University, Shenzhen, China. (E-mail: \url{sanglw21@mails.tsinghua.edu.cn}, \url{xu.yinliang@sz.tsinghua.edu.cn}.)}
  \thanks{Zhongkai~Yi is Department of Electrical Engineering, Harbin Institute of Technology, Harbin, China. (E-mail: \url{yzk_article@163.com})}
  \thanks{Lun~Yang is with the Ministry of Education Key Laboratory for Intelligent Networks and Network Security, Xi'an Jiaotong University, Xi'an, 710049, China. (E-mail: \url{yanglun2019@gmail.com})}
  \thanks{Huan~Long is with the School of Electrical Engineering, Southeast University, Nanjing, China. (E-mail:\url{hlong@seu.edu.cn})}
  \thanks{Hongbin~Sun is with the Department of Electrical Engineering, State Key Laboratory of Power Systems, Tsinghua University, Beijing, China. (E-mail: \url{shb@tsinghua.edu.cn})}
}

\begin{document}

\maketitle
{\rv 
\begin{abstract}
  The increasing penetration of distributed energy resources (DERs) will decrease the rotational inertia of the power system and further degrade the system frequency stability. To address the above issues, this paper leverages the advanced neural network (NN) to learn the frequency dynamics and incorporates NN to facilitate system reliable operation. This paper proposes the conservative sparse neural network (CSNN) embedded frequency-constrained unit commitment (FCUC) with converter-based DERs, including the learning and optimization stages. In the learning stage, it samples the inertia parameters, calculates the corresponding frequency, and characterizes the stability region of the sampled parameters using the convex hulls to ensure stability and avoid extrapolation. For conservativeness, the positive prediction error penalty is added to the loss function to prevent possible frequency requirement violation. For the sparsity, the NN topology pruning is employed to eliminate unnecessary connections for solving acceleration. In the optimization stage, the trained CSNN is transformed into mixed-integer linear constraints using the big-M method and then incorporated to establish the data-enhanced model. The case study verifies 1) the effectiveness of the proposed model in terms of high accuracy, fewer parameters, and significant solving acceleration; 2) the stable system operation against frequency violation under contingency.
\end{abstract}
}


\section*{Nomenclature}
The main notation below is provided for unit commitment problem formulation. And other additional symbols are defined in the paper when needed.
\begin{description}[leftmargin=10em,style=nextline]
  \item[A. Sets and Indices]
\end{description}
\begin{description}[leftmargin=8em,style=nextline]
  \item[$t \in \mathcal{T}$] Set of the time periods.
  \item[$g \in \mathcal{G}$] Set of the thermal generators.
  \item[$n \in \mathcal{N}$] Set of the system buses.
  \item[$w \in \mathcal{W}$] Set of the converter-based wind turbine generators.
  \item[$pv \in \mathcal{PV}$] Set of the converter-based PV generators.
  \item[$j \in \mathcal{N}_d$] Set of the droop control converters.
  \item[$k \in \mathcal{N}_v$] Set of the VSM control converters.  
\end{description}
\begin{description}[leftmargin=10em,style=nextline]
  \item[B. Decision Variables]
\end{description}
\begin{description}[leftmargin=8em,style=nextline]
  \item[$x_{g,t}^{SU}/x_{g,t}^{SD}$] Indicators of units start up / shunt down.
  \item[$x_{g, t}$] Operation status of units.
  \item[$p_{g,t}/p_{w,t}/p_{pv,t}$] Power output of thermal, wind turbines, and PV generators.
  \item[$p^{cur}_{n,t}$] Load curtailment of the bus node.
  \item[$u_{gi}/u_{dj}/u_{dk}$] Frequency response participation indicators of the thermal generators, droop / VSM converters.
  \item[$M$] System aggregate inertia constant.
  \item[$D$] System aggregate damping constant.
  \item[$R_g$] System aggregate droop factor.
  \item[$F_g$] System aggregate fraction factor.      
  \item[$\delta$] Phase angle of the bus node. 
  \item[$P_{ij}$] Power flow between node $i$ and node $j$.
\end{description}
\begin{description}[leftmargin=10em,style=nextline]
  \item[C. Parameters]
\end{description}
\begin{description}[leftmargin=8em,style=nextline]
  {\rv \item[$f_{N}$] Nominal frequency rating value. }
  {\rv \item[$f^{min}/f_{lim}(t_{j,t})$] Minimum frequency limits / stepwise frequency limit. }
  \item[$C_{oper}/C_{rsv}/C_{load}$] Costs of system operation / reserve / load reduction.
  \item[$B_{mn}$] The susceptance of the transmission line. 
  \item[$\hat{p}_{w, t}/\hat{p}_{pv, t}$] Day-ahead predicted power of wind / PV.
  \item[$M_{gi}$] Inertia constant of the unit $i$. 
  \item[$D_{gi}$] Damping constant of the unit $i$. 
  \item[$R_{gi}$] Inertia constant of the unit $i$. 
  \item[$K_{gi}$] Mechanical power gain of the unit $i$. 
  \item[$F_{gi}$] Fraction of the unit $i$ power generation by the turbine.
  \item[$R_{dj}$] Droop gain of the converter $j$.
  \item[$K_{dj}$] Mechanical power gain of the converter $j$.
  \item[$M_{dk}$] Virtual inertia of the converter $k$.
  \item[$D_{dk}$] Virtual damp constant of the converter $k$.
  {\rv\item[$\epsilon_{l}/\epsilon_{der}$] Prediction errors of load and DERs.}
\end{description}

\section{Introduction} \label{sec: intro}

\IEEEPARstart{I}{n} the transition to {\rv a} low carbon energy society, the renewable energies are replacing the conventional fossil energies and taking up a larger proportion of generation in the modern power system \cite{Chilvers2021}. Unlike the direct connection of fossil-based synchronous generators, the distributed energy resources (DERs) are integrated into the grid through power converters that lack inherent rotational inertia, leading to the low-inertia systems \cite{Milano2018}. The decreased rotational inertia will degrade the frequency stability of {\rv the system} or even trigger large-scale blackouts \cite{Delille2012, Cui2022}.

To address the above frequency stability problems of the low-inertia systems, the system operator should consider the frequency dynamics under contingency, transform the frequency dynamics requirements into static operational constraints, and incorporate them into the upstream scheduling models. Current research has focused on the formulation and incorporation of the frequency constraints by approximating the frequency nadir, which can be categorized into two types. The first type approximates the frequency response of generators as a piece-wise linear function to time, decoupling the governor control from the frequency in \refcite{Teng2016, Badesa2019, Badesa2020, Schipper2020I}. Though exhibiting a simplified form with the frequency nadir constraints, it ignores the actual control policy and may not apply to high penetrated converter-based DER scenarios. The second type is based on the system frequency response model combined with the converter response feature to formulate the total frequency dynamics in \refcite{Paturet2020, Zhang2020, Yuan2022, Zhang2022}. Ref. \cite{Paturet2020, Zhang2020} derive the analytical expression of frequency dynamics with converter-based DERs, calculate the frequency nadir expression, and utilize the piece-wise linear formulation to approximate these dynamics, which can be converted into a set of linear constraints. Ref. \cite{Zhang2022} further considers the joint scheduling of wind farm frequency support and reserve based on the analytical expression. {\rv These piece-wise linear models include the converter-based DER response but have limited approximation capability}, which may not satisfy {\rv the requirement of complex converter-based power systems} \cite{Wang2019, Auba2021}. 

{\rv The frequency support potential of the high-penetration DERs has been} explored and exploited within the control scheme of the grid-forming voltage source converter (VSC) \cite{Chen2022, Markovic2019, Yan2020}. Virtual synchronous machine policy is one of the common VSC control strategies based on the emulation of {\rv system} inertia \cite{Markovic2019}. Ref. \cite{Chen2022} proposes a generalized architecture of VSC from the view of multivariable feedback control to unify different control strategies. Ref. \cite{Yan2020} focuses on the wind turbine generators and proposes multiple virtual rotating masses for frequency support. However, the direct inertia parameter aggregation of different VSC control methods may lead to the low-inertia system instability \cite{Milano2018}.

So the relationship between the frequency dynamics and frequency support sources under the high penetrated converter-based DER is governed by the system's ordinary differential equations, which are complicated and cannot be incorporated into the system operation model directly. The thriving machine learning (ML) technique, {\rv widely applied in pattern cognition \cite{Xia2022,Zhang2022_r1_1} and image detection \cite{Chen2021,Zhang2022_r1_2},} provides the powerful deep neural networks (DNNs) to formulate the above relationship with high accuracy \cite{Maragno2021,Fajemisin2021,Bengio2021,Bertsimas2022,Sun2020}. The enforced constraints on the DNN-based frequency can be transformed into the mixed-integer linear programming \cite{Anderson2020} and solved efficiently by the off-the-shelf solvers \cite{Bertsimas2022}. {\rv In Ref. \cite{Maragno2021}, the DNN is utilized to learn constraints and objectives, and the learned DNN is embedded in solving the world food program planning problem.} |{\rv Two critical issues arises from the direct integration of the trained DNNs into} the frequency-constrained scheduling models will raise : {\rv 1) the positive and negative prediction errors of a DNN have different impacts on the system, where positive errors can lead to violations of the frequency requirements, and} 2) the parameter number of the conventional dense DNN can be huge, resulting in the long solving time \cite{Evci2020}. 

Faced with the complex frequency dynamics under the high penetrated converter-based DERs, this paper leverages the advanced research in ML and the dramatic improvement in solving mixed-integer linear programming (MILP) to formulate and incorporate the frequency constraints into operation models {\rv to ensure} the system stable operation. To address the issues of DNN integration, it proposes the \textit{conservative sparse neural network} (CSNN) embedded frequency-constrained unit commitment model with DER, composed of learning and optimization stages. Its key lies in the design of the CSNN to achieve the conservativeness and sparsity of NN, {\rv which can further facilitate the safety and efficiency of CSNN incorporation}. It first formulates and derives the highly non-convex frequency dynamics explicitly under DER. In the learning stage, it samples the inertia parameters, calculates the corresponding frequency, and characterizes the stability region of the sampled parameters using the convex hull to ensure stability and avoid extrapolation. Then for the conservativeness of NN, the positive prediction error penalty is added to the loss function to avoid possible frequency requirement violation; for the sparsity of NN, the neural network topology pruning is employed in training NN to eliminate some unnecessary connections for achieving MILP solving acceleration. In the optimization stage, the trained CSNN is transformed into a series of mixed-integer linear constraints using the big-M method to relax the ReLU activation functions in NN. {\rv Then these frequency constraints} is incorporated into the conventional unit commitment (UC) to {\rv establish} the final data-enhanced frequency constrained UC model in the MILP form. {\rv For this paper, the main contributions are as follows:}

1) To the best of authors' knowledge, this paper, \textit{for the first time}, proposes the \textit{conservative sparse neural network} to achieve both the conservativeness and sparsity of neural networks, which is beneficial to NN embedding in the optimization problems and can be utilized to formulate the relationship between the frequency dynamics and system inertia parameters. The proposed CSNN achieves higher approximation accuracy, compared with the previous piece-wise linear formulation of frequency dynamics in \refcite{Markovic2019, Zhang2020}, {\rv which} exhibits i) fewer parameters to accelerate MILP solving ii) and conservative prediction to prevent the violation of frequency requirements, compared with conventional {\rv NN}.

2) Based on the proposed CSNN, this paper proposes the data-enhanced frequency constrained unit commitment under DER to address the {\rv frequency stability problem of the low-inertia system}. The proposed scheduling model approximates both frequency nadir and stepwise constraints by CSNN, transforms the trained CSNNs into a series of mixed-integer linear constraints using the big-M method, and embeds the constraints into the conventional UC model. Compared with conservative and conventional neural networks, the proposed data-enhanced model can achieve significant MILP solution time reduction, verifying its efficiency.

3) This paper designs the \textit{stability region} using convex hulls from the sampled dataset with stability to guarantee the system operation stability and avoid extrapolation. The main reason for \textit{stability region} is that not all inertia parameters in the sampled dataset can stabilize the total system, especially under the high penetration of the converter-based DERs. So the inertia parameters with stability are extracted to construct the stable dataset, and then the clustered convex hulls are extracted, denoted by \textit{stability region}, which are transformed into linear constraints and embedded into the scheduling model to guarantee the stability and feasibility.

{\rv For clearer delivery, we further summarize the main motivation and innovation of this paper. The increasing distributed energy resources will decrease the system rotational inertia and further degrade the system frequency stability, which requires the frequency consideration in system scheduling. However, the relationships between frequency dynamics and frequency support sources are complex, which requires the accurate and conservative formulation to capture the relationship. To address the above issues, we propose the CSNN with high representational capacity for capturing the above complex frequency relationship effectively with conservativeness design and embed the trained CSNN into the conventional unit commitment model efficiently with sparsity design to formulate the data-enhanced frequency constrained unit commitment model.} The rest of this paper is organized as follows. Section \ref{sec: prob} depicts the frequency dynamics with distributed energy resources. Section \ref{sec: method} proposes the conservative sparse neural network to formulate the frequency constraints. Section \ref{sec: fcuc} presents the CSNN embedded data-enhanced frequency-constrained unit commitment with DER. Section \ref{sec: case study} performs the case study to verify the efficiency of the proposed model. Section \ref{sec: conclusion} concludes this paper.

\section{Frequency Dynamics Formulation with Distributed Energy Resources} \label{sec: prob}

This section formulates the system frequency dynamics under conventional \textit{synchronous} generators and \textit{converter-based} DER generators mathematically, governed by nonlinear ordinary differential equations from \refcite{Kundur2017}. The frequency dynamics can be analyzed by the swing equation in \eqref{eq: swing}.
\begin{eqnarray}
  \begin{aligned}
    M \frac{d \Delta f}{d t} + D \Delta f = \Delta p_g - \Delta p_e .
  \end{aligned}
  \label{eq: swing}
\end{eqnarray}
Then, based on \eqref{eq: swing}, the low-inertia system dynamics is introduced, and several key frequency constraints are proposed to guarantee the frequency dynamic stability.

\subsection{Frequency Dynamics Analytical Analysis}

Based on \refcite{Markovic2019}, we can derive the transfer function $G(s)$ of the system dynamics model in \eqref{eq: analytical transfer}. \footnote{The detailed infrastructure can be referred to \refcite{Paturet2020}.}

\begin{eqnarray}
  \allowdisplaybreaks
  \begin{aligned}
    G(s) & = \frac{\Delta f}{\Delta P_e} = \Big(\underbrace{(s M_g + D_g) + \sum_{i \in \mathcal{N}_g}\frac{K_{gi}(1+sF_{gi}T_{gi})}{R_{gi}(1+sT_{gi})}}_{\text{Multiple-machine system frequency response}} \\
    & + \underbrace{\sum_{j\in\mathcal{N}_d} \frac{K_{cj}}{R_{cj}(1+sT_{cj})}}_{\text{Droop control from DERs}}  
    + \underbrace{\sum_{k\in\mathcal{N}_v} \frac{sM_{ck}+D_{ck}}{1+sT_{ck}}}_{\text{VSM control from DERs}} \Big)^{-1}.
  \end{aligned}
  \label{eq: analytical transfer}
\end{eqnarray}

It is assumed that all synchronous generators have equal time constants with $T_{gi} = T, \forall i \in\mathcal{N}_g$. The time constants of converters are 2-3 orders lower than those of synchronous generators with $T \gg T_c \approx 0$, according to \refcite{Ahmadi2014}. Then we transform the general-order \eqref{eq: analytical transfer} into its second-order form \eqref{eq: general form}.
\begin{eqnarray}
  \allowdisplaybreaks
  \begin{aligned}
    G(s) = \frac{1}{MT}\frac{1+sT}{s^2+2\zeta\omega_n s + \omega^2_n}.
  \end{aligned}
  \label{eq: general form}
\end{eqnarray}
where $\omega_n$ and $\zeta$ denote the natural frequency and damping ratio, calculated by \eqref{eq: param cal}.
\begin{eqnarray}
  \allowdisplaybreaks
  \begin{aligned}
    \omega_n = \sqrt{\frac{D+R_g}{MT}}, \quad \zeta=\frac{M+T(D+F_g)}{2\sqrt{MT(D+R_g)}}.
  \end{aligned}
  \label{eq: param cal}
\end{eqnarray}
The key parameters $M$, $D$, $F_g$, and $R_g$ in \eqref{eq: param cal} calculation can be referred to \refcite{Markovic2019}.

Given a stepwise disturbance $\Delta P $ ($\Delta P / s$ in the frequency domain) in \eqref{eq: swing}, the frequency dynamics in time domain can be derived from \eqref{eq: general form} as follows: 
\begin{eqnarray}
  \allowdisplaybreaks
  \begin{aligned}
    \Delta f(t) = & - \frac{\Delta P}{MT \omega^2_n} \\
    &- \frac{\Delta P}{M \omega_d}e^{-\zeta\omega_n t}\Big(\sin(\omega_d t) - \frac{1}{\omega_n T}\sin(\omega_d t + \phi) \Big)
  \end{aligned}
  \label{eq: time domain}
\end{eqnarray}
where the $\omega_d$ and $\phi$ calculated as follows:
\begin{eqnarray}
  \allowdisplaybreaks
  \begin{aligned}
     \omega_d = \omega_n \sqrt{1 - \zeta^2}, \quad \phi = \arcsin(\sqrt{1-\zeta^2}).
  \end{aligned}
  \label{eq: new param def}
\end{eqnarray}

The frequency nadir is located in the time instance $t_m$ where the derivative of frequency is zero in \eqref{eq: t nadir}.
\begin{eqnarray}
  \allowdisplaybreaks
  \begin{aligned}
    \frac{d \Delta f(t_{m})}{t_{m}} = 0 \rightarrow
    t_{m} = \frac{1}{\omega_d} \tan^{-1}(\frac{\omega_d}{\zeta \omega_n - T^{-1}}).
  \end{aligned}
  \label{eq: t nadir}
\end{eqnarray}
Then the maximum frequency deviation can be calculated by replacing the $t$ of \eqref{eq: time domain} by $t_{m}$ of \eqref{eq: t nadir} as:
\begin{eqnarray}
  \allowdisplaybreaks
  \begin{aligned}
    \Delta f_{max} = -\frac{\Delta P}{D + R_g}(1 + \sqrt{\frac{T(R_g- F_g)}{M}}e^{- \zeta \omega_n t_m}).
  \end{aligned}
  \label{eq: f nadir}
\end{eqnarray}

\subsection{Frequency Constraints Formulation}

Based on the above analytical frequency response formulation, this paper proposes four key constraints to bound frequency dynamics for {\rv securing the system stability from different perspectives}, including the rate-of-change-of-frequency (RoCoF) {\rv constraint for guaranteeing transient stability \cite{Daly2019}, the quasi-steady state for bringing back system normal operation state \cite{Paturet2020}, the frequency nadir constraint for avoiding the activation of under frequency load shedding \cite{Markovic2019}, and the stepwise frequency constraint for enhancing the frequency restoration performance\cite{Schipper2020I}}.

\subsubsection{RoCoF Constraint}

RoCoF constraint guarantees the sufficient system inertia to limit the allowable RoCoF at $t=0$, following the ENTSO-E standard \cite{NG2017}.
\begin{equation}
  |\text{RoCoF}| = |\frac{\Delta P}{M}| \leq \text{RoCoF}^{max}.
  \label{eq: rocof contrs}
\end{equation}

\subsubsection{Quasi-Steady State Constraint}

Quasi-steady state (QSS) constraint guarantees quasi-steady state security assuming the RoCoF is effectively zero. 
\begin{equation}
  \Delta f_{qss} = - \frac{\Delta P}{D + R_g} \leq \Delta f^{max}_{qss}.
  \label{eq: qss contrs}
\end{equation}

\subsubsection{Frequency Nadir Constraints}

To avoid triggering under-frequency load shedding (UFLS) and oscillation, the maximum frequency deviation should be limited within the allowable range by \eqref{eq: nadir contrs}. 
\begin{equation}
  f_{nadir} = f_N - \Delta f_{max} \geq f^{min}.
  \label{eq: nadir contrs}
\end{equation}

\subsubsection{Frequency Stepwise Constraints}

\begin{figure}[ht]
  \centering
  \includegraphics[scale=1.3]{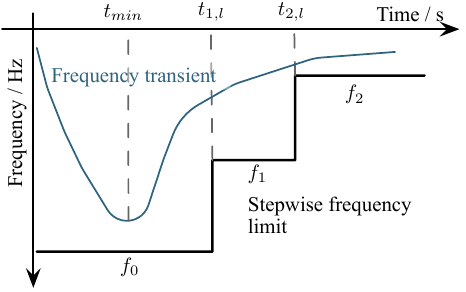}
  \caption{Multiple stepwise frequency requirement.}
  \label{fig: freq multireq}
\end{figure}

Besides the above frequency constraints, we further consider the frequency restoration requirement from frequency nadir to quasi-steady state, inspired by \refcite{Schipper2020I}. The stepwise frequency constraints are utilized to bound the frequency restoration process, as shown in \eqref{eq: mulifreq}, \eqref{eq: freq stepwise}, and Fig. \ref{fig: freq multireq}.

\begin{eqnarray}
  \allowdisplaybreaks
  \begin{aligned}
    f_{lim} (t) = 
    \begin{cases}
      f_0 \quad & 0 \leq t < t_{1, l} \\
      f_1 \quad & t_{1, l} \leq t <  t_{2, l} \\
      \cdots \\
      f_j \quad & t_{j, l} \leq t <  t_{j+1, l} \\
      \cdots \\
      f_{N_c} \quad & t_{N_c, l} \leq t 
    \end{cases}.
  \end{aligned}
  \label{eq: mulifreq}
\end{eqnarray}

Then we formulate the stepwise frequency constraints as follows: 
\begin{eqnarray}
  \allowdisplaybreaks
  \begin{aligned}
    f(t) = f_{N} - \Delta f(t) \geq f_{lim}(t).
  \end{aligned}
  \label{eq: freq stepwise}
\end{eqnarray}
However, the above constraints are complicated because they should be evaluated for $t\geq0$. As shown in Fig. \ref{fig: freq multireq}, the stepwise frequency constraints are binding when the transient curve and limit curve touch. The constraints are not required when no step changes in the frequency limit curve. So we only evaluate it at the step change time $t = t_{j,l}$ as follows:
\begin{eqnarray}
  \allowdisplaybreaks
  \begin{aligned}
    f(t_{j,l}) = f_{N} - \Delta f(t_{j,l}) \geq f_{lim}(t_{j,l}) \quad j \in N_c .
  \end{aligned}
  \label{eq: freq stepwise trans}
\end{eqnarray} 

\subsection{Frequency Constraints Analysis} \label{sec: prob C}

\subsubsection{Relationship Analysis}

The RoCoF and QSS constraints are linear with respect to $M$, $D$, and $R_g$. In contrast, the frequency nadir and stepwise constraints are nonlinear due to the nonlinearity of $f_{nadir}(M, D, R_g, F_g)$ and $f(M, D, R_g, F_g; t_{j,l})$ with respect to $M$, $D$, $R_g$, and $F_g$.

\subsubsection{Conservative Approximation of Frequency Constraints}

So we need to derive the tractable constraints from the nonlinear ones to facilitate tractable solving. The system operator prioritizes the negative error (predicted value minus true value) but is averse to positive error, where positive errors may result in the frequency requirement violation. So we propose to approximate the $f_{nadir}$ and $f(t_{j,l})$ in a conservative way in \eqref{eq: freq conser}. 
\begin{subequations}
  \allowdisplaybreaks
  \begin{align}
    & f_{nadir}(M, D, R_g, F_g) \geq \hat{f}_{nadir}(M, D, R_g, F_g) \geq f^{min} \label{eq: freq conser a}\\ 
    & f(M, D, R_g, F_g; t_{j,l}) \geq \hat{f}(M, D, R_g, F_g; t_{j,l}) \geq f_{lim}(t_{j,l}) \label{eq: freq conser b}
  \end{align}
  \label{eq: freq conser}
\end{subequations}
where $\hat{f}_{nadir}$ and $\hat{f}$ are the predicted frequency, which are the {\rv lower} bounds of the true frequency.

\section{Methodology} \label{sec: method}

Based on the frequency constraints formulation and analysis, this section 1) introduces the concept of optimization with constraint learning (OCL) framework, 2) proposes the conservative sparse neural network formulation under the OCL framework for nonlinear frequency constraints in \eqref{eq: freq conser}, 3) embeds the trained neural network into the optimization models via MILP, 4) and presents two data processing techniques for facilitating predictive model integration.

{\rv
\subsection{Data-driven Optimization with Constraint Learning Framework Design}
}

Practical optimization problems often contain constraints with intractable forms or no explicit formulas. However, based on the supportive data corresponding to such constraints, these data can be utilized to learn the constraints (constraint learning, CL). The trained predictive models can be embedded into the optimization models. So the above two stages are called optimization with constraint learning (OCL). We denote the final tractable model by the data-enhanced optimization model. 

According to the review of \refcite{Fajemisin2021}, the OCL framework is usually composed of the following five steps: 1) formulate the conceptual optimization model 2) conduct data sampling and processing; 3) construct and train predictive models; 4) embed the predictive model into the optimization model; 5) solve and evaluate the data-enhanced optimization model. It should be noted that the OCL framework can be extended to objective learning by transforming the objective function into constraints via {\rv the epigraph of the function} \cite{Maragno2021}.

\subsection{Conservative Sparse Neural Network Formulation}

Under the above OCL framework, we propose the CSNN to learn the relationship of \eqref{eq: freq conser} and formulate the data-enhanced frequency-constrained unit commitment model based on the trained CSNN, detailed in the following parts. {\rv The} learning of \eqref{eq: freq conser} is denoted by frequency nadir constraint learning for \eqref{eq: freq conser a} and stepwise constraint learning for \eqref{eq: freq conser b}. {\rv The proposed CSNN prediction models are both conservative and sparse, which can guarantee the safety of frequency constraints and accelerate the NN-embedding model solution.} 

\subsubsection{Conservative Neural Network Design}

Based on the analysis of frequency constraints in section \ref{sec: prob C}, the negative errors are prioritized against possible frequency requirement violations compared to the positive errors. So the conservative prediction models are prioritized for the system {\rv safe and stable} operation. To this end, we propose a conservative loss function where prediction values are always lower than the true values. It is calculated by the weighted sum of \textit{mean-square-error} loss function and additional positive error ReLU loss function to penalize the positive part of the frequency constraints, as illustrated in \eqref{eq: conservative loss}. 

\begin{subequations}
  \small
  \begin{gather}
    \hat{f}^*_\theta \leftarrow  \arg\min_{\theta} \mathcal{L}^{com}_{\theta} \label{eq: com f} \\
    \mathcal{L}^{com}_{\theta} = \underbrace{\frac{1}{2}\mathbb{E}_{(x,y)\sim \mathcal{D}} (\hat{f}_\theta (x) - y)^2}_{\text{supervised regression}}
    + \underbrace{\mathbb{E}_{(x,y)\sim \mathcal{D}}\text{ReLU}(\hat{f}_\theta (x) - y) }_{\text{ReLU loss function}} \label{eq: com loss}
  \end{gather}
  \label{eq: conservative loss}
\end{subequations}
where the resulting $\hat{f}^*_\theta(x)$ is the conservative estimate of the actual function $f(x)$.

\subsubsection{Sparse Neural Network Design}

\begin{figure}[ht]
  \centering
  \includegraphics[scale=1]{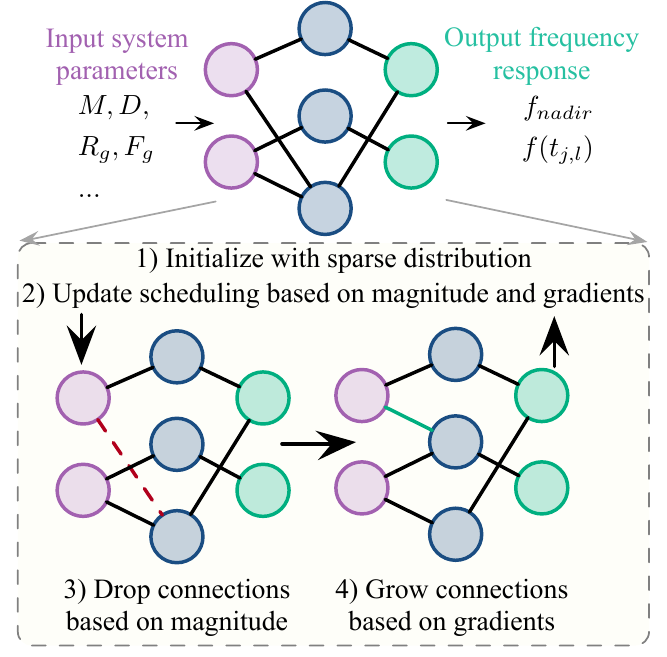}
  \caption{Sparse neural network training by leveraging the weight magnitude and gradient information.}
  \label{fig: sparse scheme}
\end{figure}

We propose to train sparse neural networks by dropping and activating their inner connections iteratively to alleviate the calculation burden of dense full-connected NN {\rv embedding}. The keys to sparse training lie in updating neural network topology based on parameter magnitude criterion for dropping connections and infrequent gradient criterion for activating connections, which is from the lottery ticket hypothesis in \refcite{Frankle2019}. {\rv The motivation behind the growing and dropping sparse training is as follows: the dropping of neuron connections may sacrifice the accuracy of the learning models; so, we further re-activate $k$ connections with the top-$k$ high magnitude gradients, because high magnitude gradients indicate high impact connection between neurons. So we re-activate these connections for improving the accuracy.} The main scheme of sparse training is composed of sparse distributing, update scheduling, connection dropping, and connection activating for learning the mapping from the input system parameter to the output frequency response, as shown in Fig. \ref{fig: sparse scheme}.

\paragraph{Sparse Distributing}

We define sparsity $s^l \in (0, 1)$ as the dropping fraction of the $l^{th}$ layer. The first layer is set as the dense layer, and the other layers are set as the sparse layers with predefined $s^l$ for the $l^{th}$ layer sparse training in a uniform way. 

\paragraph{Update Scheduling}

The update scheduling drops a fraction of connections based on NN parameter magnitude and activates new ones based on NN parameter gradient magnitude iteratively at $\Delta T$ interval. So the updating fraction function by considering decaying is defined in \eqref{eq: frac up}.
\begin{eqnarray}
  \begin{aligned}
    f_{decay}(t; \alpha, T_{end}) = \frac{\alpha}{2} (1+\cos \frac{t\pi}{T_{end}})
  \end{aligned}
  \label{eq: frac up}
\end{eqnarray}
where $T_{end}$ is the number of iteration for sparse training, and $\alpha$ is the initial drop fraction.

\paragraph{Connection Dropping}

In connection dropping of the $l^{th}$ layer, we utilize $ArgTopK(v,k)$ algorithm to select the top-$k$ maximum element indices of vector $v$ and drop the connection by $ArgTopK(-|\theta^l|, f_{decay}(t;\alpha, T_{end})(1-s^l)N^l)$.

\paragraph{Connection Activating}

After dropping, we further re-activate $k$ connections with the top-$k$ high magnitude gradients by $ArgTopK(\nabla_{\theta^l}L^{com}_t , f_{decay}(t;\alpha, T_{end})(1-s^l)N^l)$. Newly activated connections are initialized as zero to avoid influencing the network output.

{\rv
\begin{remark}
  The effect of gradients on the model's training lies in the following two aspects: {i)} the gradients from the loss function can learn the prediction models by minimizing the proposed conservative loss function ({stochastic gradient training}); {ii)} they can also be utilized to grow necessary connection in the  {sparse neural networks} to improve accuracy.
\end{remark}
}

\subsubsection{Conservative Sparse SGD Algorithm}

Based on the conservative loss of \eqref{eq: com loss} and sparse training designs, we propose the conservative sparse stochastic gradient descent (CS-SGD) algorithm for training the CSNN in algorithm \ref{algo: cs-sgd}.

\begin{algorithm}[ht]
  {\small{\rv
    \begin{algorithmic}[1]
      \STATE \textbf{Input:} {\rv Initialize the frequency dynamics model:} $f_{\theta}$, {\rv dataset: $D = \{x_i, y_i\}$,} sparse distributing: $\mathbb{S}=\{s^1, .., s^L\}$, updating scheduling {\rv parameters}: $\Delta T$, $T_{end}$, $\alpha$, $f_{decay}$;
      \STATE {\rv \textbf{Output:} Trained frequency dynamics model $f_{\theta}$.}
      \FOR {$t=1,...,T$}
      \STATE {\textbf{Conservative Formulating:}}
      \STATE Calculate conservative loss $L^{com}_t = \sum L^{com}(f_{\theta}(x_i), y_i)$ based on \eqref{eq: conservative loss};
      \STATE {\textbf{Sparse Training:}}
      \IF {$t \bmod \Delta T == 0$ and $t \leq T_{end}$}
      \FOR {$l= 1,...,L$}
      \STATE Calculate dropping and activating number \\ $k = f_{decay}(t;\alpha, T_{end})(1-s^l)N^l$;
      \STATE Drop {\rv connections based on} $ArgTopK(-|\theta^l|, k)$;
      \STATE Activate {\rv connections based on} $ArgTopK(\nabla_{\theta^l}L^{com}_t , k)$;
      \ENDFOR
      \ELSE
      \STATE {\rv \textbf{Update CSNN parameters: }}$\theta \leftarrow \theta - \alpha \Delta_{\theta} L^{c om}_t$;
      \ENDIF
      \ENDFOR
    \end{algorithmic}}
  }
  \caption{Conservative Sparse Stochastic Gradient Descent Training Algorithm.}
  \label{algo: cs-sgd}
\end{algorithm}

{\rv The key novelty of the proposed CS-SGD algorithm lies in incorporating the conservative loss function of \eqref{eq: conservative loss} and sparse training process with dropping and growing to train CSNN with conservativeness and sparsity for approximating the power system frequency dynamics. The weights and bias of trained CSNN are leveraged to formulate the equivalent mixed-integer linear frequency constraints, which can secure the system frequency stability. The conservativeness of CSNN guarantees safety of the transformed frequency constraints, and the sparsity of CSNN improving solution efficiency of the data-enhanced frequency constrained unit commitment optimization model with fewer optimizing variables.
}

\subsection{NN Embedded Frequency Constraints Transformation}

After training, the proposed CSNN can be transformed into a series of mixed linear integer constraints using the big-M method. Based on \refcite{Anderson2020}, {\rv this} paper formulates the general form of an $L$-layer feedforward neural network with ReLU as the activation function in \eqref{eq: nn form}, which is composed of stacked linear transformation (\ref{eq: nn form a}) and ReLU activation (\ref{eq: nn form b}).
\begin{subequations}
  \allowdisplaybreaks
  \begin{align}
    \text{Neural Network:} &v^l_i = \text{ReLU} (\theta^l_{i0} + \sum_{j\in\mathcal{N}^{l-1}} \theta^l_{ij} v^{l-1}_j) \label{eq: nn form a} \\
    \text{where  } & y = v^L = \theta^L_{0} + \sum_{j\in\mathcal{N}^{L-1}} \theta^L_{j} v^{L-1}_j \label{eq: nn form b}
  \end{align}
  \label{eq: nn form}
\end{subequations}
{\rv where $v^l_i$ is the output of neuron $i$ in layer $l$ in a neuron network, $\theta^l_{ij}$ is the weight between neuron $i$ and $j$, and $\theta^l_{i0}$ is the bias of neuron $i$ in layer $l$.}

For each layer of \eqref{eq: nn form a}, the ReLU operator, $\text{ReLU}(x)=\max\{0, x\}$, can be encoded using linear constraints in a big-M way from \refcite{Maragno2021}, as follows:

\begin{subequations}
  \allowdisplaybreaks
  \begin{align}
    v & \geq x, \\
    v & \leq x - M_L(1-t), \\
    v & \leq M_U t, \\
    v & \geq 0, \\
    t & \in \{0, 1\}
  \end{align}
  \label{eq: big-m}
\end{subequations}
where $M_U$ and $M_L$ are positive and negative large value to relax the ReLU function.

\subsection{Data Processing Techniques}

This part utilizes the data sampling strategy based on the boundary and Latin hypercube sampling to enhance the data quality for the predictive models. It constructs the stability region based on the convex hulls to prevent the extrapolating of predictive models.

\subsubsection{Data Sampling Strategy}

The predictive model performance relies on the training dataset quality; thus, the data distribution is critical for constraint learning. The sampled data should be space-filling sufficiently to capture the frequency dynamics of frequency constraints in \eqref{eq: freq conser}. Inspired by Ref. \cite{Bertsimas2022}, we utilize the boundary sampling and optimal Latin hypercube (OLH) sampling for constructing training set $\mathcal{Z} = \{M_{n}, D_{n}, R_{g, n}, F_{g,n}\}_{n=1}^{N}$. In the process, the boundary sampling collects the corners of the $\mathcal{Z}$ hypercube by the input data bounds, and then we conduct the OLH sampling over the above hypercube. The detailed sampling procedures are illustrated in \refcite{Bertsimas2022}.

\subsubsection{Convex Hulls as Stability Region}

The optimal solutions of mixed-integer linear programs are often located at the extremes of their feasible polyhedron region, which may harm the validity of the trained ML models. According to \refcite{Maragno2021}, the accuracy of an ML-based predictive model will deteriorate for points far away from the training dataset. So we construct stability regions to prevent the predictive model from extrapolating by the convex hull from the training dataset. For the training dataset $\mathcal{Z}$, we construct the following convex hull $CH(\mathcal{Z})$ in \eqref{eq: ch z} as the stability region.

\begin{eqnarray}
  \begin{gathered}
    CH(\mathcal{Z}) = \big\{ z | \sum_{i\in \mathcal{I}} \lambda_i \hat{z}_i = z, \sum_{i\in\mathcal{I}} \lambda_i = 1, \lambda \geq 0 \big\}
  \end{gathered}
  \label{eq: ch z}
\end{eqnarray}

However, there are two issues arising from the direct single convex hull. One issue is that the computing of the convex hull is exponential to data number in time and space; the other is that the data distribution in a single convex hull is not uniform. Some points cannot satisfy the requirement of system stability in \eqref{eq: analytical transfer}, and the single convex hull may include and extrapolate some infeasible points. So we propose the cluster-then-build strategy, which first clusters the trained dataset, builds a convex hull for each cluster, and unions them $\bigcup_{k\in\mathcal{K}} CH_{k}(\mathcal{Z})$ for preventing extrapolating in \eqref{eq: K ch z}.
\begin{eqnarray}
  \begin{aligned}
    \bigcup_{k\in\mathcal{K}} CH_{k}(\mathcal{Z}) = \big\{ z | \sum_{i\in \mathcal{I_k}} \lambda_i \hat{z}_i = z, \sum_{i\in\mathcal{I}} \lambda_i = u_k, \sum_{k\in\mathcal{K}}u_k=1 \\ 
    \lambda \geq 0, \forall k, u \in \{0,1\} \big\}
  \end{aligned}
  \label{eq: K ch z}
\end{eqnarray}

{\rv 
\subsubsection{Discussion}

It should be noted that we need to construct and build different CSNN models and stability regions for different system and DER settings. Because the training of CSNNs is based on the sampled dataset from the system model with corresponding DER setting via \eqref{eq: time domain} and \eqref{eq: f nadir}. Different system and DER settings will affect the sampled dataset directly and further affect the training of CSNNs indirectly, leading to inaccurate approximation. Stability region is also constructed from the sampled dataset. So we need to retrain different learning models and construct different stability region for these circumstances, such as different DER settings, to improve approximation accuracy.
}

\section{Data-enhanced Frequency-Constrained Unit Commitment with DER} \label{sec: fcuc}

{\rv Based on the above proposed CSNN,} this section proposes the data-enhanced frequency constrained unit commitment (FCUC) with the converter-based DER. It firstly introduces the overall OCL framework for {\rv the proposed data-enhanced} FCUC, then presents the conventional unit commitment model, and includes the data-enhanced frequency constraints to formulate the final data-enhanced model.

\subsection{Overall Framework}

\begin{figure}[ht]
  \centering
  \includegraphics[scale=1.2]{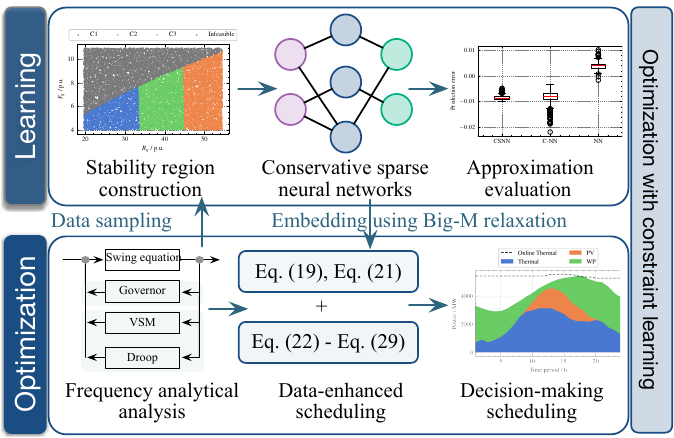}
  \caption{The optimization with constraint learning framework for the data-enhanced FCUC with DER.}
  \label{fig: ocl workflow}
\end{figure}

Based on the CSNN and NN embedded transformation of section \ref{sec: method}, we propose the OCL framework for the data-enhanced FCUC with converter-based DERs, composed of the learning and optimization stages, as shown in Fig. \ref{fig: ocl workflow}. Its main idea lies in leveraging CSNN and {\rv the frequency} data to enhance the power system decision-making and {\rv advance} the data-driven integration modeling technique. The learning stage samples the stable instances from the frequency analytical expression of \eqref{eq: f nadir} and \eqref{eq: freq stepwise}, {\rv feeds them to training the CSNNs, constructs the stability region using the convex hulls for the stable instances,} and evaluates their performance; the optimization stage embeds the stability region constraints of \eqref{eq: K ch z} and transformed CSNN constraints of \eqref{eq: big-m} in the system scheduling model to formulate the final data-enhanced FCUC model. {\rv It should be noted that: 1) the stability region and the CSNNs are constructed from the sampled dataset in different ways and are integrated into the conventional unit commitment model as mixed-integer linear constraints, so the stability region construction will not affect the training process of CSNNs; 2) DERs refer to the distributed photovoltaic (PV) and wind power (WP) generators concretely in this paper.}

\subsection{Unit Commitment Considering Joint Reserve and Regulation Capacity}

Based on the above OCL framework, we formulate the conventional unit commitment model considering joint reserve and regulation capacity to minimize the system operating cost under a series of operational constraints.

\subsubsection{Model Objective}

The objective of conventional unit commitment is to minimize the total system operating cost, {\rv which is composed of the} operating {\rv cost} $C_{oper}$, reserve {\rv cost} $C_{rsv}$, and load shedding {\rv cost} $C_{load}$ in \eqref{eq: uc obj}.
\begin{subequations}
  \allowdisplaybreaks
  \begin{align}
    &\min \quad C_{total} = C_{oper} + C_{rsv} + C_{load} \\ 
    &C_{oper} = \sum_{t\in\mathcal{N}_T}\sum_{g\in\mathcal{N}_G}(C^{SU}_g x^{SU}_{g,t} + C^{SD}_g x^{SD}_{g,t} + f_g(p_{g,t})) \\
    & C_{rsv} = \sum_{t\in\mathcal{N}_T} \Big(\sum_{g\in\mathcal{G}} C^{R}_{g} R_{g,t} + \sum_{w\in\mathcal{W}} C^{R}_{w} R_{w,t} + \sum_{pv\in\mathcal{PV}} C^{R}_{pv} R_{pv,t} \Big) \\
    & C_{load} =\sum_{t\in\mathcal{N}_T}\sum_{n\in\mathcal{N}_N} \lambda_L p^{cur}_{n,t}
  \end{align}
  \label{eq: uc obj}
\end{subequations}
where $C_{oper}$ includes the start-up,  shut-down, and the fuel costs; $C_{rsv}$ includes the costs of the reserve from synchronous generators and converter-based {\rv PV/WP} generators; $C_{load}$ includes the load shedding costs from different buses. The unit cost function $f_g(\cdot)$ can be piece-wised by SOS2 constraints.

\subsubsection{Power Network Constraints}

The node power balance equation and transmission power limited are formulated {\rv based on DC power flow model} in \eqref{eq: uc pnc}.

\begin{subequations}
  \allowdisplaybreaks
  \begin{align}
    &\sum_{g\in\mathcal{G}_n}p_{g,t} + \sum_{w\in\mathcal{W}_n}p_{w,t} + \sum_{pv\in\mathcal{PV}_n}p_{pv, t} \nonumber \\
    &+ \sum_{m:(n,m)}B_{nm}(\delta_{nt}-\delta_{mt}) = p^D_{n,t}- p^{cur}_{n,t} \quad \forall n, t \label{eq: uc pnc a} \\ 
    & |B_{nm}(\delta_{nt}-\delta_{mt})| \leq F^{max}_{nm} \quad \forall (n,m)\in\mathcal{L}, \forall t \label{eq: uc pnc b}
  \end{align}
  \label{eq: uc pnc}
\end{subequations}
where \eqref{eq: uc pnc a} enforces the bus power balance for each period, and \eqref{eq: uc pnc b} restricts the branch power flow under the given capacity.

\subsubsection{Thermal Generator Operational Constraints}

The ramping and power output restriction are formulated via MILP in \eqref{eq: uc opera}.

\begin{subequations}
  \allowdisplaybreaks
  \begin{align}
    x_{g,t} p^{min}_g \leq p_{g,t} \leq x_{g,t} p^{max}_g \quad \forall g, t \label{eq: uc opera a}\\
    x^{SU}_{g,t} - x^{SD}_{g,t} = x_{g,t} - x_{g, t-1} \quad \forall g, t \label{eq: uc opera b}\\ 
    p_{g,t} - p_{g,t-1}\leq x_{g,t-1} Ru_{g} \quad \forall g, t  \label{eq: uc opera c}\\ 
    p_{g,t-1} - p_{g,t}\leq x_{g,t} Rd_{g} \quad \forall g, t \label{eq: uc opera d}
  \end{align}
  \label{eq: uc opera}
\end{subequations}
where \eqref{eq: uc opera a} restricts the generator power output; \eqref{eq: uc opera b} enforces the relationship between the on/off variables and start-up/shut-down variables; \eqref{eq: uc opera c} and \eqref{eq: uc opera d} restricts the generator ramping capacity.

\subsubsection{Renewable Energy Generation Constraints}

The actual power outputs of renewable energy generation are limited below the predicted values.

\begin{subequations}
  \allowdisplaybreaks
  \begin{align}
    0 &\leq p_{w, t} \leq \hat{p}_{w, t} \quad \forall w, t \label{eq: uc res a}\\ 
    0 &\leq p_{pv, t} \leq \hat{p}_{pv, t} \quad \forall pv, t \label{eq: uc res b}
  \end{align}
  \label{eq: uc res}
\end{subequations}
where \eqref{eq: uc res a} and \eqref{eq: uc res b} restrict the actual power output of wind and PV below the day ahead predicted value.

\subsubsection{System Reserve Constraints}

The system reserve is restricted to cover the prediction errors of the DERs and load  as follows:
\begin{eqnarray}
  \allowdisplaybreaks
  \begin{aligned}
    &\sum_{g\in\mathcal{G}}x_{g,t}P^{max}_{g} + \sum_{w\in\mathcal{W}}\hat{p}_{w,t} + \sum_{pv\in\mathcal{PV}}\hat{p}_{pv,t} + \sum_{n\in\mathcal{N}} p^{cur}_{n,t} \\
    &\geq (1+\epsilon_{l})\sum_{n\in\mathcal{N}} \hat{p}^{D}_{n,t} \\ 
    &+ (1+\epsilon_{der})\big(\sum_{w\in\mathcal{W}}\hat{p}_{w,t} + \sum_{pv\in\mathcal{PV}}\hat{p}_{pv,t} \big) \quad \forall t .
  \end{aligned}
  \label{eq: uc rsv}
\end{eqnarray}
\eqref{eq: uc rsv} guarantees that the online generation capacity can satisfy the load demand with prediction errors, according to Ref. \cite{Zhang2020}.

\subsection{Data-enhanced Frequency Constraints}

As discussed in section \ref{sec: method}, we propose the CSNN to {\rv approximate} the nonlinear constraints of frequency nadir and stepwise constraints in a conservative way. So this part adds the CSNN embedded frequency dynamics constraints in \eqref{eq: uc freq dyn} and corresponding emergency regulation constraints in \eqref{eq: uc em rsv} into the above conventional unit commitment models for ensuring the frequency dynamics stability.

\subsubsection{Inertia Parameter Aggregation}

The system inertia variables $M$, $D$, $R_g$, and $F_g$ for \eqref{eq: freq conser} can be calculated by aggregating generators' and PV/WP converters' related parameters in \eqref{eq: inertia agg}.
\begin{subequations}
  \allowdisplaybreaks
  \begin{align}
    & M_g = \frac{\sum_{i\in\mathcal{N}_g} u_{gi} M_{gi}P_{gi}}{P^{b}_g},  D_g = \frac{\sum_{i\in\mathcal{N}_g} u_{gi} D_{gi}P_{gi}}{P^{b}_g} \\
    & F_{g} = \sum_{i\in\mathcal{N}_g} \frac{u_{gi}K_{gi}F_{gi}}{R_{gi}}\frac{P_{gi}}{P^{b}_g}, R_{g}= \sum_{i\in\mathcal{N}_g} \frac{u_{gi}K_{gi}}{R_{gi}}\frac{P_{gi}}{P^{b}_g} \\ 
    & R_{d} = \sum_{j\in\mathcal{N}_{d}} \frac{u_{dj}K_{dj}}{R_{dj}}\frac{P_{dj}}{P^{b}_{der}} \\ 
    & M_{d} = \sum_{k\in\mathcal{N}_{v}} u_{dk} M_{dk}\frac{P_{dk}}{P^{b}_{der}}, D_{d} = \sum_{k\in\mathcal{N}_{v}} u_{dk} D_{dk}\frac{P_{dk}}{P^{b}_{der}} \\ 
    & M = \frac{M_g P^{b}_g + M_{der}P^{b}_{der}}{P^{b}_g + P^{b}_{der}} \\ 
    & D = \frac{D_g P^{b}_g + D_{der}P^{b}_{der} + R_{d}P^{b}_{der}}{P^{b}_g + P^{b}_{der}} .
  \end{align}
  \label{eq: inertia agg}
\end{subequations}

\subsubsection{Frequency Dynamics Constraints}

The proposed CSNN is utilized to reach the convex {\rv lower} bounds of nonlinear frequency constraints by transforming $f_{nadir}$ and $f(t_{j,l})$ into $\hat{f}^{csnn}_{nadir}$ and $\hat{f}^{csnn}(t_{j,l})$ in \eqref{eq: nadir contrs} and \eqref{eq: freq stepwise trans}, as follows:

\begin{subequations}
  \allowdisplaybreaks
  \begin{align}
    & \hat{f}^{csnn}_{nadir}(M, D, R_g, F_g) \geq f_{min} & & \label{eq: uc freq dyn a}\\ 
    & \hat{f}^{csnn}(M, D, R_g, F_g; t_{j,l}) \geq f_{lim}(t_{j,l}). \label{eq: uc freq dyn b}
  \end{align}
  \label{eq: uc freq dyn}
\end{subequations}
where $\hat{f}^{csnn}_{nadir}$ and $\hat{f}^{csnn}(t_{j,l})$ can be transformed into mixed-integer linear constraints via \eqref{eq: big-m}. {\rv Taking \eqref{eq: uc freq dyn a} as an example, its equivalent mixed-integer constraints is formulated and incorporated by taking $x = (M, D, R_{g}, F_{g})$ in \eqref{eq: freq milp}.

\begin{subequations}
  \allowdisplaybreaks
  \begin{align}
    &v^{1}_{i} = \theta^1_{i0} + \sum_{j\in\mathcal{X}} \theta^l_{ij} x_j, & & i \in [N_1] \\ 
    &\hat{v}^l_i = \theta^l_{i0} + \sum_{j\in\mathcal{N}^{l-1}} \theta^l_{ij} v^{l-1}_j, & &l\in [L], i\in[N_l] \\
    &{v}^l_i \geq \hat{v}^l_i, & &l\in [L], i\in[N_l] \\ 
    &{v}^l_i \leq \hat{v}^l_i - M^{L}_{l,i} (1-t_{l,i}), & &l\in [L], i\in[N_l] \\ 
    &{v}^l_i \leq M^{H}_{l,i} t_{l,i}, & &l\in [L], i\in[N_l] \\ 
    &{v}^l_i \geq 0, \quad t_{l,i} \in \{0, 1\}, & &l\in [L], i\in[N_l] \\ 
    &\hat{f}^{csnn}_{nadir, m} = \theta^L_{i0} + \sum_{j\in\mathcal{N}^{L-1}} \theta^L_{ij} v^{L-1}_j & &m\in[N_M], i\in[N_L]
  \end{align}
  \label{eq: freq milp}
\end{subequations}
where $[N_l]$ is the set of input neurons in layer $l$, and $[L]$ is the set of output neurons in layer $l$.
}
\subsubsection{Emergency Regulation Formulation}
\eqref{eq: uc em rsv} provides the headroom regulation capacity for the primary frequency response from synchronous and converter-based generators.
\begin{subequations}
  \allowdisplaybreaks
  \begin{align}
    &R_{g,t} = p^{max}_g - p_{g,t} \geq u_{g,t}\frac{K_g}{R_{g}} p^{max}_g \gamma (f_0 - f_{min}) \label{eq: uc em rsv a}\\ 
    &R_{w,t} = \hat{p}^{w,t} - p_{w,t} \geq u_{w,t}\frac{1}{R_{w}} \hat{p}^{w,t} \gamma (f_0 - f_{min}) \label{eq: uc em rsv b}\\
    &R_{pv,t} = \hat{p}^{pv,t} - p_{pv,t} \geq u_{pv,t}\frac{1}{R_{pv}} \hat{p}^{pv,t} \gamma (f_0 - f_{min}) \label{eq: uc em rsv c}\\
    & \forall g\in\mathcal{G}, w\in\mathcal{W}, pv\in\mathcal{PV}, \quad \forall t \nonumber .
  \end{align}
  \label{eq: uc em rsv}
\end{subequations}
The largest frequency deviation ($f_0 - f_{min}$) is larger the quasi-steady state frequency deviation. According to \refcite{Zhang2020}, the $\gamma$ is set as 0.5 for emergency reserve capacity.

So the combination of unit commitment modeling and data-enhanced frequency constraints {\rv constructs} the {\rv final} data-enhanced FCUC with converter-based DER considering both the system reserve for uncertainty and the regulation capacity for emergency, which {\rv follows the mixed-integer linear programming form} and can be solved by the off-the-shelf solvers efficiently.

\section{Case Study} \label{sec: case study}

In the case study, the modified IEEE {\rv 118-bus} and {\rv 1888-bus} systems with multiple PV and WP is utilized to verify the effectiveness and efficiency of the data-enhanced frequency constrained unit commitment model with converter-based DER. The proposed model is composed of the learning and optimization stages. So we construct the learning {\rv models} with the Pytorch package and the optimization {\rv models} with the Cvxpy package equipped by {\rv the Gurobi solver}. All the experiments are implemented in a MacbookPro laptop with RAM of 16 GB, CPU Intel Core I7 (2.6 GHz). The other detailed system parameter setting of the modified system is attached in Ref. \cite{system_param} {\rv and Ref.~\cite{Xavier2022}}.

{\rv
\subsection{Data Preparation  and Experiment Setting}
}

This part mainly prepares the essential data and problem settings in the OCL framework for the data-enhanced FCUC {\rv model} of Fig. \ref{fig: ocl workflow}{\rv,} including 1) the neural network parameter setting of the learning stage 2) and the inertia parameter {\rv setting} of the optimization stage. 

\subsubsection{Learning Stage}

{\rv In the learning stage, we first generate 10000 feasible samples based on the system frequency dynamics of \eqref{eq: time domain} and \eqref{eq: f nadir} to construct the training dataset for the stability region and learning model training. Then the proposed CSNNs take the system inertia parameter $x = (M, D, R_{g}, F_{g})$ as the feature vector, and the input feature vector is normalized into the scale of 0-1 for model training based on their maximum value.} We compare the proposed conservative sparse neural network (CSNN) with the conservative neural network (C-NN) and the conventional neural network (NN) under the same initial parameters of full connected neural networks in Tab.~\ref{tab: nn param}. {\rv The hyperparameter configuration in Tab.~\ref{tab: nn param} is determined based on the hyperparameter tuning with the high approximation accuracy of learning models. The change of these hyperparameter values does not enhance the approximation accuracy of the learning models in our cases.}

\begin{table}[ht]
  \renewcommand{\arraystretch}{1.3}
  \centering
  \caption{The parameters of neural network models.}
  \begin{tabular}{cc|cc}
    \hline
    Hyperparameter & Value & Hyperparameter & Value \\
    \hline
    Optimizer & Adam & Learning rate & 1e-6 \\ 
    Hidden layers & [100, 100] & Batch size & 32 \\ 
    Activation function & ReLU & Dropout & 0.2 \\
    \hline
  \end{tabular}
  \label{tab: nn param}
\end{table}

We clarify the relationship among CSNN, C-NN, and NN, where C-NN further utilizes the conservative loss of \eqref{eq: conservative loss} compared to NN, and CSNN further employs the sparse training of algorithm \ref{algo: cs-sgd} compared to C-NN. {\rv The same hyperparameters of Tab.~\ref{tab: nn param} are utilized to initialize the CSNN, C-NN, and NN to avoid the hyperparameter effect on model comparison of different neural networks.}

\subsubsection{Optimization Stage}

In the optimization stage, we prepare the inertia parameters of different generators and VSM/droop control parameters of PV and wind power converters from \refcite{Markovic2019}, as illustrated in Tab. \ref{tab: inertia param}. The frequency nadir, stepwise frequency, and QSS requirement are set based on the ENTSO-E standard \cite{NG2017} and Ref. \cite{Schipper2020I}.

\begin{table}[ht]
  \renewcommand{\arraystretch}{1.3}
  \centering
  \caption{The parameters of different generator and control methods.}
  \begin{tabular}{cccccc}
    \hline
    Type & $M_{gi}$ & $K_{gi}$ & $F_{gi}$ & $R_{ci}$ & $D_{ci}$ \\
    \hline
    Nuclear & 9s & 0.98 & 0.25 & 0.04 & 0.6 \\
    CCGT & 14s & 1.1 & 0.15 & 0.01 & 0.6 \\
    OCGT & 11s & 0.95 & 0.35 & 0.03 & 0.6 \\
    \hline
    Converter / VSM & 12s & 1.0 & - & - & 0.6 \\
    Converter / Droop & - &1.0 & - & 0.05 & - \\
    \hline
  \end{tabular}
  \label{tab: inertia param}
\end{table}

\subsection{Frequency Approximation Performance Evaluation}

This part firstly analyzes the approximation performance of the proposed CSNN in describing the frequency dynamics. It then verifies its effectiveness in the data-enhanced frequency constrained model with the stability region design.

\subsubsection{CSNN Analysis}

\paragraph{Training Process}

\begin{figure}[ht]
  \centering
  \subfigure[Training process of the frequency nadir.]{\includegraphics[scale=0.95]{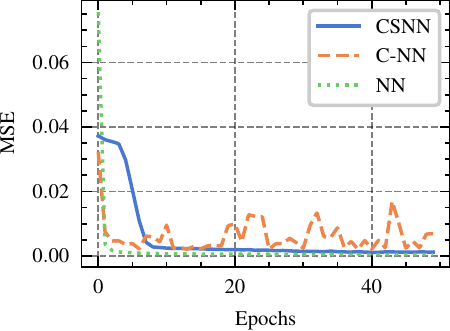}}
  \subfigure[Training process of the stepwise frequency constraints.]{\includegraphics[scale=0.95]{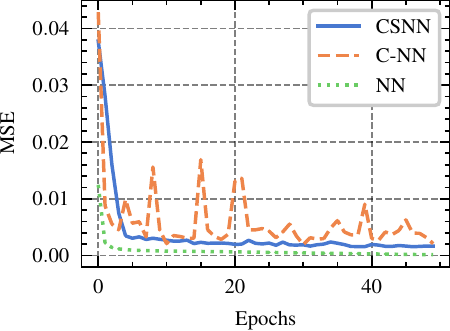}}
  \caption{{\rv The training process of frequency nadir and stepwise frequency constraint learning.}}
  \label{fig: training}
\end{figure}

Fig. \ref{fig: training} compares CSNN, C-NN, and NN training processes for the frequency nadir and stepwise relationship approximation. In the training process, NN achieves a lower approximation error than CSNN and C-NN due to the consideration of the conservative loss function, and CSNN exhibits a little lower error than C-NN thanks to the sparse designs.

\paragraph{Approximation Error}

\begin{figure}[ht]
  \centering
  \includegraphics[scale=1]{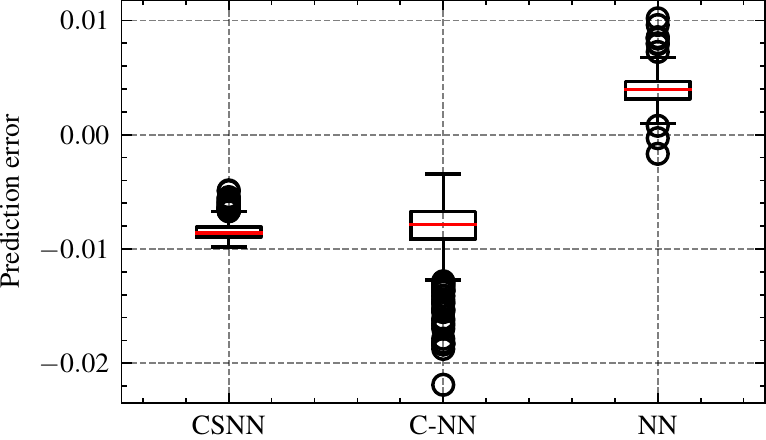}
  \caption{{\rv The approximation errors of different models in boxen plot.}}
  \label{fig: pred error}
\end{figure}

We further compare the detailed approximation performance of CSNN with C-NN and NN. Fig. \ref{fig: pred error} demonstrates the approximation error distribution of the above four models by the boxen plot.

As shown in Fig. \ref{fig: pred error}, the approximation errors of NN are located around zero with negative and positive values, and in contrast, the approximation errors of CSNN and C-NN are located below zero with only negative values, demonstrating their conservativeness. The median error of NN is above zero. CSNN exhibits a more accurate approximation with a more concentrated error distribution than C-NN.

We utilize the mean square error (MSE), mean absolute percentage error (MAPE) metrics in \eqref{eq: mse mape}, and negative rate to measure the approximation performance numerically.

\begin{eqnarray}
  \begin{aligned}
    \mathcal{L}_{MSE} = \frac{1}{N} \sum_{i}^{N} (f_i - \hat{f}_i)^2, \quad \mathcal{L}_{MAPE} = \frac{1}{N} \sum_{i}^{N} |\frac{f_i - \hat{f}_i}{f_i}|.
  \end{aligned}
  \label{eq: mse mape}
\end{eqnarray}

As illustrated in Tab. \ref{tab: pred error}, for MAPE, CSNN achieves 0.87\% lower MAPE than NN, 0.76\% lower MAPE than C-NN, and 0.11\% lower MAPE than PWL; for MSE, CSNN increases 142.14\% MSE of NN but decrease 91.35\% MSE of C-NN and 74.85\% MSE of C-NN, which sacrifices some approximation accuracy to guarantee conservativeness. As for the negative rate, both CSNN and C-NN exhibit a 100\% negative rate, demonstrating the effectiveness of the conservative loss design. In contrast, the negative rate of NN is only 85.43\%, which may result in the frequency requirement violation.

\begin{table}[ht]
  \renewcommand{\arraystretch}{1.3}
  \centering
  \caption{Approximation error comparison of different methods.}
  \begin{tabular}{ccccc}
    \hline
    Models & CSNN & C-NN & NN & PWL \\
    \hline
    MAPE & 0.23\% & 1.10\% & 0.99\% & 0.44\%  \\ 
    MSE & $7.584 e^{-5}$& $8.772 e^{-4}$& $3.132 e^{-5} $ & $3.015 e^{-4}$ \\ 
    Negativeness & 100\% & 100\% & 85.43\% & 65.46\% \\
    \hline
  \end{tabular}
  \vspace{1ex}
  
  {\raggedright We note that 1) the negativeness denotes the ratio of the negative error instance number to the total number; 2) PWL denotes the conventional piece-wise linear approximation. \par}
  \label{tab: pred error}
\end{table}

\paragraph{Sparsity Comparison}

To verify the sparsity of CSNN, we compare the non-zero parameter number of the above four models, where lower parameters mean a higher sparsity.

\begin{table}[ht]
  \renewcommand{\arraystretch}{1.3}
  \centering
  \caption{The parameter number of different methods for frequency nadir.}
  \begin{tabular}{ccccc}
    \hline
    Models & CSNN & C-NN & NN \\
    \hline
    Number & 1395 & 10701 & 10701 \\ 
    Sparsity rate & 86.96\% & 0\% & 0\% \\
    \hline
  \end{tabular}
  \vspace{1ex}
  
  {\raggedright We note that the sparsity rate denotes the ratio of the parameter reduction number to the total parameter number. \par}
  \label{tab: param nadir num}
\end{table}

For the frequency nadir constraint learning of \eqref{eq: freq conser a} in Tab. \ref{tab: param nadir num}, the corresponding CSNN has only 1395 parameters and achieves the 86.96\% reduction of the dense neural networks of C-NN and NN, verifying the effectiveness of sparse training.

\begin{table}[ht]
  \renewcommand{\arraystretch}{1.3}
  \centering
  \caption{The parameter number of different methods for stepwise constraints.}
  \begin{tabular}{ccccc}
    \hline
    Models & CSNN & C-NN & NN \\
    \hline
    Number & 1442 & 10903 & 10903 \\
    Sparsity rate & 86.77\% & 0\% & 0\% \\
    \hline
  \end{tabular}
  \label{tab: param sw num}
\end{table}

For the stepwise frequency constraint learning of \eqref{eq: freq conser b} in Tab. \ref{tab: param sw num}, the corresponding CSNN has only 1442 parameters and achieves the 86.77\% reduction of the dense neural networks of C-NN and NN.

{\rv
\subsection{Comparison of Optimization Scenarios}
}

In this part, we first compare the operational performance of the data-enhanced frequency constrained unit commitment model and then verify the {\rv efficacy} of the sparse design of neural networks, stability region, and the stepwise frequency constraints.

\subsubsection{Operation Performance}

We further compare the data-enhanced FCUC with DER to the conventional unit commitment for verifying the necessity of frequency constraints and the effectiveness of DER participation.

\paragraph{Results}

Tab. \ref{tab: oper cost} compares the operational costs and renewable abandoned rates of the scheduling model with/without frequency constraints.

\begin{table}[ht]
  \renewcommand{\arraystretch}{1.3}
  \centering
  \caption{Operational comparison of different models {\rv in the modified 118-bus system}.}
  \begin{tabular}{cccc}
    \hline
    Models & Cost / \$ & Abandoned Rate & Violation Rate \\
    \hline
    Conventional UC & $19.65e^{5}$ & 0.00\% & 100\% \\ 
    Data-enhanced FCUC & $21.27e^{5}$ & 2.64\% & 0\% \\
    \hline
  \end{tabular}
  \vspace{1ex}
  
  {\raggedright We note that the abandoned rate denotes the ratio of the abandoned renewable power to the total renewable generation. \par}
  \label{tab: oper cost}
\end{table}
The data-enhanced FCUC increases the operational cost by 8.24\% and the renewable abandoned rate by 2.64\% but decreases the system frequency violation rate from 100\% to 0\% under large contingency, verifying the effectiveness of the proposed model. The increase of abandoned renewable supports system inertia. Through the frequency constraints, the system will enhance the system inertia and frequency response by increasing the online units and exploiting the frequency support from DER, leading to a higher operational cost, abandoned rate, and more stable system operation.

\begin{figure}[ht]
  \centering
  \subfigure[Conventional scheduling.]{\includegraphics[scale=0.75]{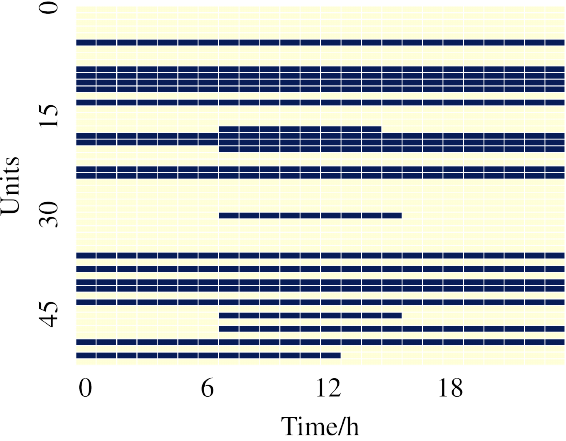}}
  \subfigure[Frequency-constrained scheduling]{\includegraphics[scale=0.75]{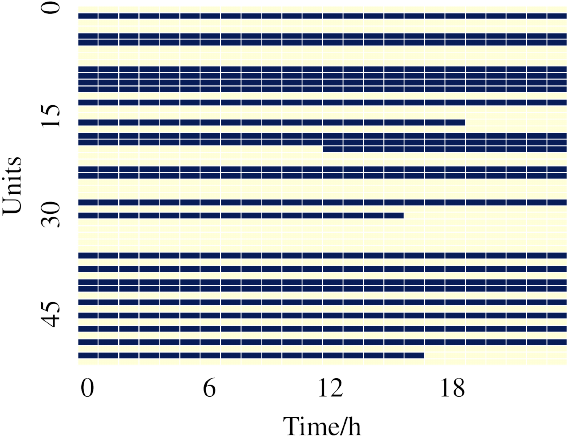}}
  \caption{\rv The unit status under different scheduling scenarios in the modified 118-bus system.}
  \label{fig: status}
\end{figure}

Fig. \ref{fig: status} further compares the operational unit status between the convention scheduling and frequency-constrained scheduling, demonstrating that the frequency constraints enforce the unit start-up to guarantee the system stability.

\paragraph{Dynamics Analysis}

\begin{figure}[ht]
  \centering
  \includegraphics[scale=1]{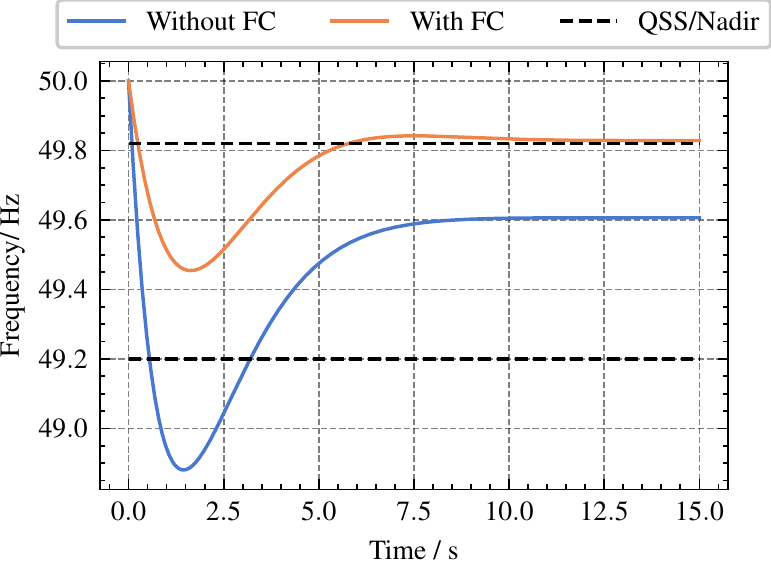}
  \caption{{\rv Frequency dynamics comparison in the modified 118-bus system.}}
  \label{fig: freq dyna comp}
\end{figure}

We simulate the frequency dynamics under the scheduling models with and without frequency constraints under the 500 MW power loss contingency in Fig. \ref{fig: freq dyna comp}, verifying their effectiveness. 

As shown in Fig. \ref{fig: freq dyna comp}, the orange line of dynamics under FCUC satisfies the frequency nadir, stepwise, and QSS requirements compared to the blue line under the conventional scheduling.

\subsubsection{Benefits of Sparse Neural Networks}

Compared with the piece-wise linear regression, the high representational capacity of sparse neural networks will help capture the relationship between frequency dynamics and inertia parameters in a more optimization-efficient way.

\begin{table}[ht]
  \renewcommand{\arraystretch}{1.3}
  \centering
  \caption{Comparison of different frequency models {\rv in the modified 118-bus system}.}
  \begin{tabular}{cccc}
    \hline
    Models & CSNN & C-NN & NN \\
    \hline
    Solving time/min & 7.12 & 122.65 & 153.55 \\ 
    \hline
  \end{tabular}
  \label{tab: snn benefit}
\end{table}

Tab. \ref{tab: snn benefit} compares the solving time and costs of the data-enhanced FCUC integrated with different neural networks. The solving time under CSNN is 17.23 times faster than C-NN and 21.57 times faster than NN, verifying the solution acceleration effectiveness of sparsity design.  

\subsubsection{Benefits of Stability Region}

\begin{figure}[ht]
  \centering
  \includegraphics[scale=1]{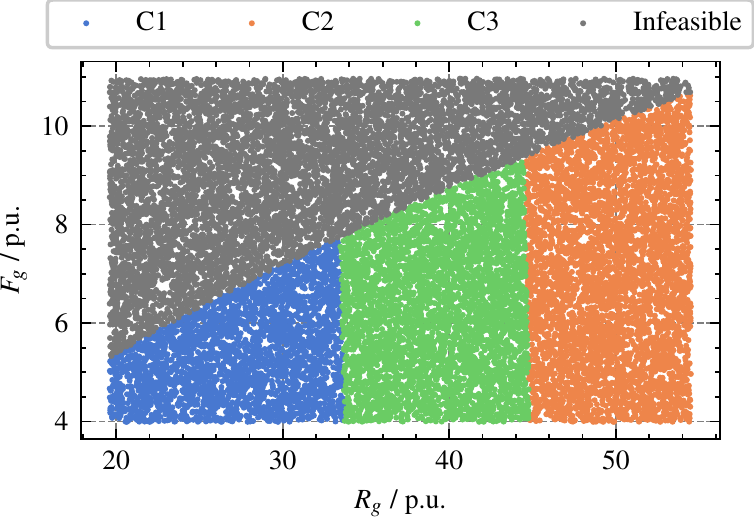}
  \caption{{\rv Analysis of convex hulls as stability region.}}
  \label{fig: ch analysis}
\end{figure}

We should guarantee the performance of the data-driven method design because robustness matters more than optimality. So we construct the clustered convex hulls from the data to build the stability region adding to the optimization models. The below Fig. \ref{fig: ch analysis} shows the area of infeasible points and the convex hulls of clustered feasible points.

{\rv
\subsection{Scalability and Comparison Analysis}

For the generality of the proposed data-enhanced FCUC, we further discuss the scalability of the proposed framework via the modified 1888-bus system experiments based on Ref.~\cite{Xavier2022} and verify the effectiveness and efficiency of the proposed CSNN via the comparison the state-of-the-art machine learning algorithms \cite{Bertsimas2021,Aprilla2021,Zhang2023}. 

\subsubsection{Scalability Analysis}

We apply the data-enhanced FCUC model in a modified 1888-bus system to verify the scalability of the proposed method that the proposed data-enhanced FCUC increases 3.6\% operational cost and 4.23\% renewable abandoned rate to achieve the 0\% frequency violation rate under contingency and ensure system stable operation, as shown in the following Tab.~\ref{tab: oper cost case scalability}.
\begin{table}[ht]
  \renewcommand{\arraystretch}{1.3}
  \centering
  \setcounter{table}{7}
  \caption{\rv Operational comparison of different models in the modified 1888-bus system.}
  \begin{tabular}{cccc}
    \hline
    Models & Cost / \$ & Abandoned Rate & Violation Rate \\
    \hline
    Conventional UC & $1.469e^{7}$ & 0\% & 100\%  \\ 
    Data-enhanced FCUC & $1.524e^{7}$ & 4.23\% & 0\%  \\
    \hline
  \end{tabular}
  \label{tab: oper cost case scalability}
\end{table}

% The results shown in Fig.~\ref{fig: freq dyna scalability} further verify the frequency stability of the proposed data-enhanced FCUC with frequency constraints in a large system, compared to convention UC without frequency constraints, demonstrating its scalability and reliability.
% \begin{figure}[ht]
%   \centering
%   \includegraphics[scale=1]{figs/comp_freq_dyna_r12.pdf}
%   \caption{{\rv Frequency dynamics comparison in 1888-bus system.}}
%   \label{fig: freq dyna scalability}
% \end{figure}

\subsubsection{Comparison Analysis}

To verify the efficiency and effectiveness, we further compare the proposed CSNN model with other state-of-the-art machine learning models including the optimal regression trees (ORT) \cite{Bertsimas2021}, random forest (RF) \cite{Aprilla2021}, and support vector regression (SVR) \cite{Zhang2023} models. 

\begin{table}[ht]
  \renewcommand{\arraystretch}{1.3}
  \centering
  \caption{\rv Approximation error comparison of different ML models in the modified 1888-bus system.}
  \begin{tabular}{ccccc}
    \hline
    Models & CSNN & ORT & RF & SVR \\
    \hline
    MAPE & 0.23\% & 0.11\% & 0.32\% &  0.35\% \\ 
    MSE & $7.584 e^{-5}$& $2.83 e^{-5}$ & $1.15 e^{-4}$ & $1.47 e^{-4}$ \\ 
    Negativeness & 100\% & 45.71\% & 55.07\% & 84.52\% \\
    \hline
  \end{tabular}
  \vspace{1ex}
  
  {\raggedright We note that the negativeness denotes the ratio of the negative error instance number to the total number.\par}
  \label{tab: pred error compare}
\end{table}
As shown in Tab.~\ref{tab: pred error compare}, though the prediction accuracy of CSNN is lower than ORT but higher than RL and SVR, the proposed CSNN achieve 100\% negative rate compared to ORT, RF, and SVR, verifying its prediction conservativeness.

To summarize, our proposed data-enhanced FCUC model exhibits scalability for various testing systems and effectiveness for approximation conservativeness compared to various ML models.
}

\section{Conclusion} \label{sec: conclusion}

{\rv Faced with the complex frequency dynamics under the high penetration of the converter-based DER, this paper proposes to learn the frequency dynamics by CSNN and incorporate CSNN into the system optimization to facilitate system reliable operation, composed of learning and optimization stages. In the learning stage, the proposed CSNN can achieve the conservativeness against frequency violation and the sparsity for solving acceleration; in the optimization stage, the trained CSNNs are transformed into MILP constraints using the big-M method and embedded into the conventional UC model to establish the final data-enhanced FCUC model. The case study verifies 1) the effectiveness of the proposed model by achieving higher approximation accuracy, around 87\% fewer parameters, and 21.57 times faster solving speed compared to that of the conventional NN in the modified 118-bus system; 2) the stable system operation against the frequency requirement violation under contingency. The proposed data-enhanced FCUC model leverages and incorporates the CSNN to manage the system inertia with DERs effectively and boost the system reliability.}

\ifCLASSOPTIONcaptionsoff
  \newpage
\fi

\begin{IEEEbiographynophoto}{Linwei Sang}(S'20) received the M.S. degree from the School of Electric Engineering, Southeast University, Nanjing, China in 2021. 
  
  He is currently pursuing the Ph.D. degree in the Tsinghua-Berkeley Shenzhen Institute, Tsinghua University, Shenzhen, China. His research includes the integration of machine learning and optimization in smart grid, the control of the distributed energy, and demand side resource management. 
\end{IEEEbiographynophoto}

\begin{IEEEbiographynophoto}{Yinliang Xu}(SM'19) received the B.S. and M.S. degrees in control science and engineering from the Harbin Institute of Technology, Harbin, China, in 2007 and 2009, respectively, and the Ph.D. degree in electrical and computer engineering from New Mexico State University, Las Cruces, NM, USA, in 2013.
  
  He is currently an Associate Professor with Tsinghua-Berkeley Shenzhen Institute, Tsinghua Shenzhen International Graduate School, Tsinghua University, Beijing, China. His research interests include distributed control and optimization of power systems, renewable energy integration, and microgrid modeling and control. 
\end{IEEEbiographynophoto}

\begin{IEEEbiographynophoto}{Zhongkai Yi}(Member, IEEE) received the B.S. and M.S. degrees in Electrical Engineering from Harbin Institute of Technology in 2016 and 2018, respectively, and the Ph.D. degree in Electrical Engineering from Tsinghua University in Jan. 2022.
  He is currently an associate professor with the Department of Electrical Engineering, Harbin Institute of Technology. From January 2022 to April 2023, He is an algorithm expert with the Machine Intelligence Research Sector, Alibaba DAMO Academy. His research interests include the optimization and machine learning in power systems.
\end{IEEEbiographynophoto}

\begin{IEEEbiographynophoto}{Lun Yang} received the Ph.D. degree in Electrical Engineering from Tsinghua University, Beijing, China, in October, 2022. He is currently an Assistant Professor in the School of Automation Science and Engineering of Xi'an Jiaotong University. Before that, he was a visiting scholar with The Chinese University of Hong Kong, Shenzhen, from October 2022 to April 2023. His research interests include power and energy systems operations and data-driven optimization under uncertainty. He was the recipient of the best paper award at 2020 IEEE PES General Meeting.
\end{IEEEbiographynophoto}

\begin{IEEEbiographynophoto}{Huan Long}(M'15)
  received the B.Eng. degree from Huazhong University of Science and Technology, Wuhan, China, in 2013, and the Ph.D. degree from the City University of Hong Kong, Hong Kong, in 2017. 
  
  She is currently an Associate Professor with the School of Electrical Engineering, Southeast University, Nanjing, China. Her research fields include artificial intelligence applied in modeling, optimizing, monitoring the renewable energy system and power system. 
\end{IEEEbiographynophoto}

\begin{IEEEbiographynophoto}{Hongbin Sun}(Fellow, IEEE) received his double B.S. degrees from Tsinghua University in 1992, the Ph.D. from Dept. of E.E., Tsinghua University in 1996. 
  
  He is now ChangJiang Scholar Chair professor and the director of energy management and control research center, Tsinghua University. He also serves as the editor of the IEEE TSG, associate editor of IET RPG, and member of the Editorial Board of four international journals and several Chinese journals. His technical areas include electric power system operation and control with specific interests on the Energy Management System, Automatic Voltage Control, and Energy System Integration.
\end{IEEEbiographynophoto}

\end{document}